\documentclass[twocolumn,prb,showkeys,superscriptaddress]{revtex4}
\usepackage{graphicx}
\usepackage[cmex10]{amsmath}
\usepackage[utf8]{inputenc}
\usepackage{float}
\usepackage{amssymb}
\usepackage[T1]{fontenc}
\usepackage{hyperref}
\begin{document}

\title{Simultaneous readout of two adjacent bit tracks with a spin-torque oscillator}

\author{Jakub Ch\k{e}ciński}

\affiliation{AGH University of Science and Technology, Department of Electronics, Al. Mickiewicza 30, 30-059 Krak\'{o}w, Poland}
\affiliation{AGH University of Science and Technology, Faculty of Physics and Applied Computer Science, Al. Mickiewicza 30, 30-059 Krak\'{o}w, Poland}

\author{Marek Frankowski}

\affiliation{AGH University of Science and Technology, Department of Electronics, Al. Mickiewicza 30, 30-059 Krak\'{o}w, Poland}

\author{Tomasz Stobiecki}
\affiliation{AGH University of Science and Technology, Department of Electronics, Al. Mickiewicza 30, 30-059 Krak\'{o}w, Poland}
\affiliation{AGH University of Science and Technology, Faculty of Physics and Applied Computer Science, Al. Mickiewicza 30, 30-059 Krak\'{o}w, Poland}

\date{\today}

\begin {abstract}

We propose a novel setup for a spin-torque oscillator reader in magnetic hard disk drive  technology. Two adjacent bit tracks are to be read simultaneously, leading to high data transfer rate and increased resilience to noise as the lateral size of the oscillator device is allowed to remain larger than the bit width. We perform micromagnetic simulations of an example system and find that the magnetization response has a clear unimodal character, which enables for detection of two bit values at the same time. We analyze the frequency of the device under the influence of two different external fields and conduct a simulation of a successful dynamic readout. We estimate the signal linewidth and signal-to-noise ratios of the setup and show that it may be potentially beneficial for magnetic readout applications.

\end{abstract}

\maketitle

\section{Introduction}

Spin-torque oscillators (STOs) are a class of devices which has attracted considerable attention because of their possible applications in microwave technologies, magnonics or magnetic field detection for magnetic recording purposes \cite{tsoi1998excitation,rippard2005injection,kiselev2003microwave,houssameddine2007spin,
deac2008bias,krivorotov2007large,kima2012spin,dumas2014recent,skowronski2012zero,zeng2013ultralow,madami2011direct,demidov2010direct,
kudo2009measurement,braganca2010nanoscale,sato2012simulations,suto2014nanoscale,checinski2017spin}. 
In the context of hard disk drive (HDD) reading head, various advantages of an STO device has been pointed out, including high data-transfer rate \cite{mizushima2010signal,sato2012simulations,kanao2016effects}, good scalability \cite{kanao2016effects} and ability to selectively read signal from multilayer media \cite{suto2014nanoscale}. However, the development of STO readers has been hindered by their potential vulnerability to magnetic noise, which becomes more problematic as the size of the device decreases, leading to smaller free layer volumes and therefore bigger spectrum linewidths\cite{braganca2010nanoscale,nagasawa2016response,chao2017scaling}. From the noise perspective, it would be beneficial for STO devices to remain relatively large. On the other hand, increasing HDD data storage areal density necessitates usage of readout devices with better spatial resolution. In fact, the lateral size of an STO reader has been often implicitly assumed to directly correspond to the size of the recording bit \cite{sato2012simulations,nagasawa2016response,checinski2017spin}. We propose a novel readout configuration which could allow such a device to read two parallel bit tracks simultaneously, leading to higher data-transfer rate and decreased magnetic noise levels, since the size of the reader can be now kept considerably larger than the size of the media bit. We perform micromagnetic simulations of an STO reader detecting two magnetic field values simultaneously and show that the magnetization response remains unimodal and that it can be possible to easily discern between all four possible states. We present a simulation of a dynamic readout of a random two-bit sequence utilizing an injection lock technique. Finally, we analyze linewidth and signal-to-noise (SNR) ratio of STO working in uniform or non-uniform fields and show that the setup we propose may be highly beneficial in terms of data transfer rate or decrease of the magnetic noise, which can prove crucial for STO applications in magnetic recording.

\section{Micromagnetic simulations of a two-bit reader}

\begin{figure}
\centering
\includegraphics[trim=0cm 4cm 0cm 0cm, width=8.5cm]{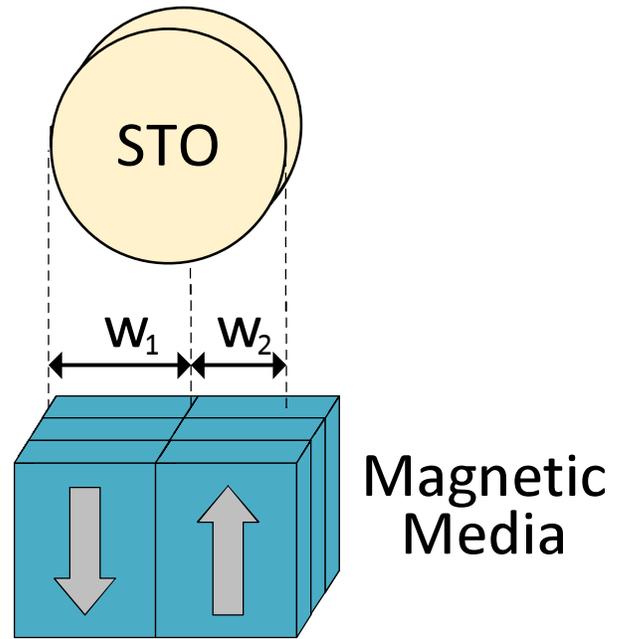}
\caption{Simplified scheme of the considered readout setup.}
\label{fig1}
\end{figure}

A simplified scheme of the readout setup we consider is presented in Figure 1. The STO is assumed to move along a magnetic media track and its lateral size $w_1 + w_2$ is assumed to be larger than the width of a single track $w_1$, but at the same time smaller than the width of two adjacent tracks combined, so that the condition $w_1 + w_2 < 2w_1$ is satisfied. For simplicity, we neglect the presence of gaps between tracks. Such positioning of the STO reader should allow fields from both bit tracks to interact with the free layer magnetization and differentiation between "01" and "10" two-bit states should in principle be possible thanks to the asymmetry condition $w_1 > w_2$. However, the character of the STO magnetization response to an explicitly non-uniform field originating from bits representing opposite values is not clear, given that most theoretical and experimental research concerning the STO dynamics have so far assumed a spatially uniform field. While a few works have already investigated the consequences of field non-uniformity in certain specific scenarios \cite{li2016micromagnetic,checinski2017spin,xia2017micromagnetic}, they tended to focus on stray field inherent inhomogeneities as a mostly detrimental effect. On the contrary, we propose to take advantage of the inhomogeneity introduced by the readout setup (Fig. 1) to detect multiple bit values simultaneously. In this section, we will describe the micromagnetic simulations we performed in order to investigate such a possibility. 

\subsection{STO frequency oscillation}

We analyzed a magnetic tunnel junction system based on our previous work \cite{checinski2017spin}, which consisted of a synthetic-antiferromagnet (SAF) structure, a non-ferromagnetic barrier of 1 nm thickness and a free layer (FL) of 2 nm thickness. The FL was assumed to be of circular shape with diameter equal to 50 nm and its saturation magnetization was set to 1400 kA/m. A small uniaxial in-plane anisotropy of 2 kJ$\mathrm{/m^3}$ was introduced in the FL along the axis orthogonal to the reference layer easy axis, similarly to the concept presented in \cite{yin2014adjusting}. The dipolar field $H_{dip}$ produced by the SAF structure in the FL volume was not equal to zero, but had a small constant value of $\mu_0H_{dip}$ = 12.1 mT, which allowed the device to work stably under both negative and positive external fields. The exchange constant was set to 1.3 $\mathrm{\times}$ 10$\mathrm{^{-11}}$ J/m and the discretization cell size was set to 2 $\mathrm{\times}$ 2 $\mathrm{\times}$ 1 nm$\mathrm{^3}$. To perform micromagnetic simulations, we utilized OOMMF package \cite{donahue1999oommf} with our custom software \cite{frankowski2014micromagnetic,chkecinski2016mage}. The magnetization dynamics was governed by the Landau-Lifshitz-Gilbert equation with Slonczewski's term \cite{landau1935theory,gilbert1955lagrangian,gilbert2004phenomenological,slonczewski1996current}, where the input current was assumed to originate from a constant voltage source and the magnetoresistance was calculated dynamically in each simulation cell \cite{frankowski2014micromagnetic}. A detailed description of the simulated system dynamics, including Arnold tongue and injection lock behavior, is provided in our previous work\cite{checinski2017spin}. 

\begin{figure}
\centering
\includegraphics[width=8.8cm]{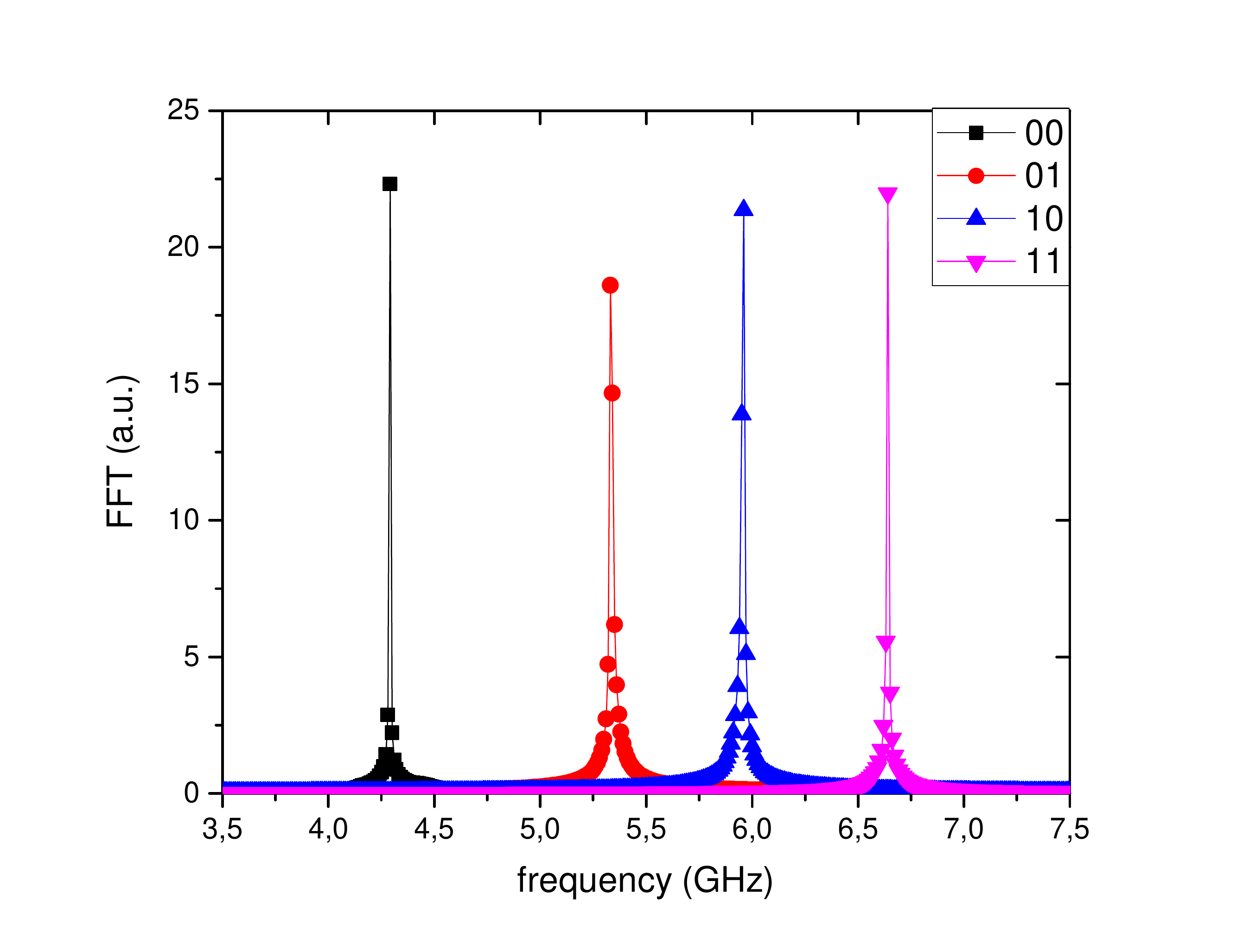}
\caption{FFT spectra presenting STO response frequency for all four two-bit states.}
\label{fig1}
\end{figure}

In order to investigate the STO response to a two-bits scenario depicted in Fig.1., we assumed $w_1$ equal to 30 nm and $w_2$ equal to 20 nm, with total lateral size of the device 50 nm. Because of circular geometry of the system, the ratio between left and right volume parts was equal to approximately 63:37 as opposed to 30:20. For simplicity, stray field value decreasing with the increasing distance from the media surface was neglected and both external fields were assumed to be spatially uniform in their respective $w_1$,$w_2$ areas. We considered a two-bit setup with both field amplitudes set to $\pm$ 5 mT, which resulted in four total scenarios corresponding to the following bit pair values: "11" (both fields set to +5 mT, average field in FL +5 mT), "10" (field in the left part +5 mT, field in the right part -5 mT, average field in FL +1.3 mT), "01" (field in the left part -5 mT, field in the right part +5 mT, average field in FL -1.3 mT) and "00" (both fields set to -5 mT, average field in FL -5 mT). The fast Fourier transform (FFT) of the device magnetization response to magnetic fields corresponding to all four possible two-bit states is presented in Figure 2. One can see that the magnetization oscillation retains a clear unimodal character even when the device is subjected to two different magnetic fields originating from bits with opposite values. Because of the spatial asymmetry of the setup, "01" and "10" scenarios are easily discernible. Moreover, neither the signal amplitude nor the response linewidth showed a significant detriment in "01" and "10" scenarios when compared to unifom "00" and "11" scenarios.  A more detailed analysis of the linewidth behavior, including influence of temperature effects, will be provided in section III. By careful manipulation of $w_1$ and $w_2$ ratio, as well as other STO parameters, further optimizations in terms of frequency change from one state to another are also possible.
\\
 \begin{figure}
\centering
\includegraphics[width=8.8cm]{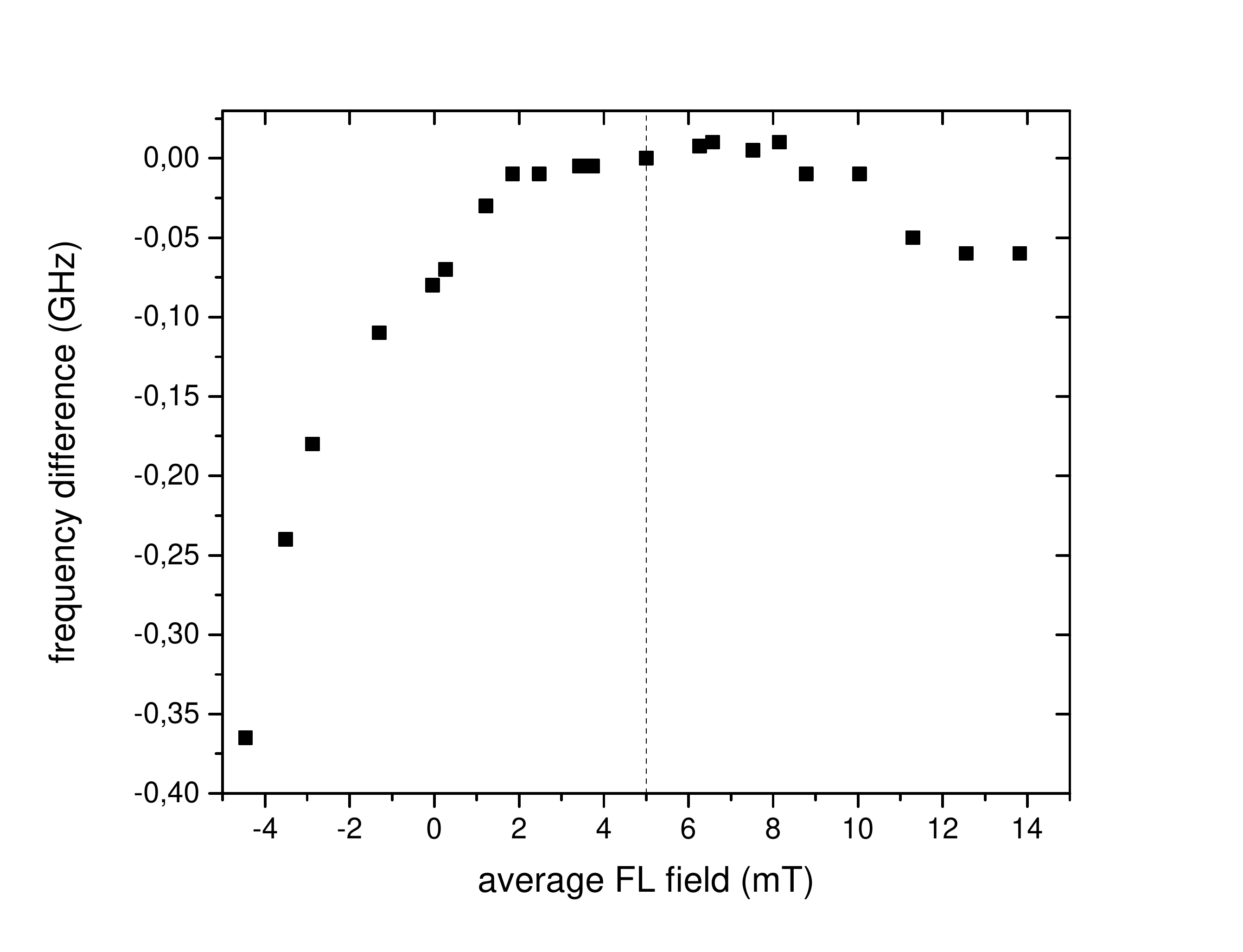}
\caption{Frequency difference between the uniform field scenario and the setup shown in Fig. 1 as a function of average field in the free layer. The dashed line indicates the $H_1 = H_2$ = +5 mT data point where uniform and non-uniform cases are indistinguishable.}
\label{fig1}
\end{figure}

Apart from the specific usage scenarios described above, it is possible to analyze the STO frequency in a non-uniform magnetic field in general. Figure 3 presents the STO frequency deviation from the uniform case as a function of the average FL field obtained from a set of micromagnetic simulations. Here, we assume $w_1$ = 30 nm, $w_2$ = 20 nm, the field applied to the right part $H_2$ is always equal to +5 mT while the field applied to the left part $H_1$ is swept in order to obtain different average fields ($H_{average} = 0.63\cdot H_1 + 0.37\cdot H_2)$. The special case of $H_{average} = H_1 = H_2$ = +5 mT, where the non-uniform field case is reduced to the +5 mT uniform field case, is indicated by the dashed line (Fig. 3). One can see that the absolute value of frequency deviation has an approximately parabolic character, increasing for larger degrees of field non-uniformity (note negative scale in Fig. 3.). However, for the proposed usage scenarios where $H_{average}$ is equal to +1.3 mT or -1.3 mT, the difference remains relatively small and does not exceed 0.1 GHz.

\subsection{Hard disk media readout simulation}

 \begin{figure}
\centering
\includegraphics[trim=1.5cm 1cm 0cm 0cm,width=9cm]{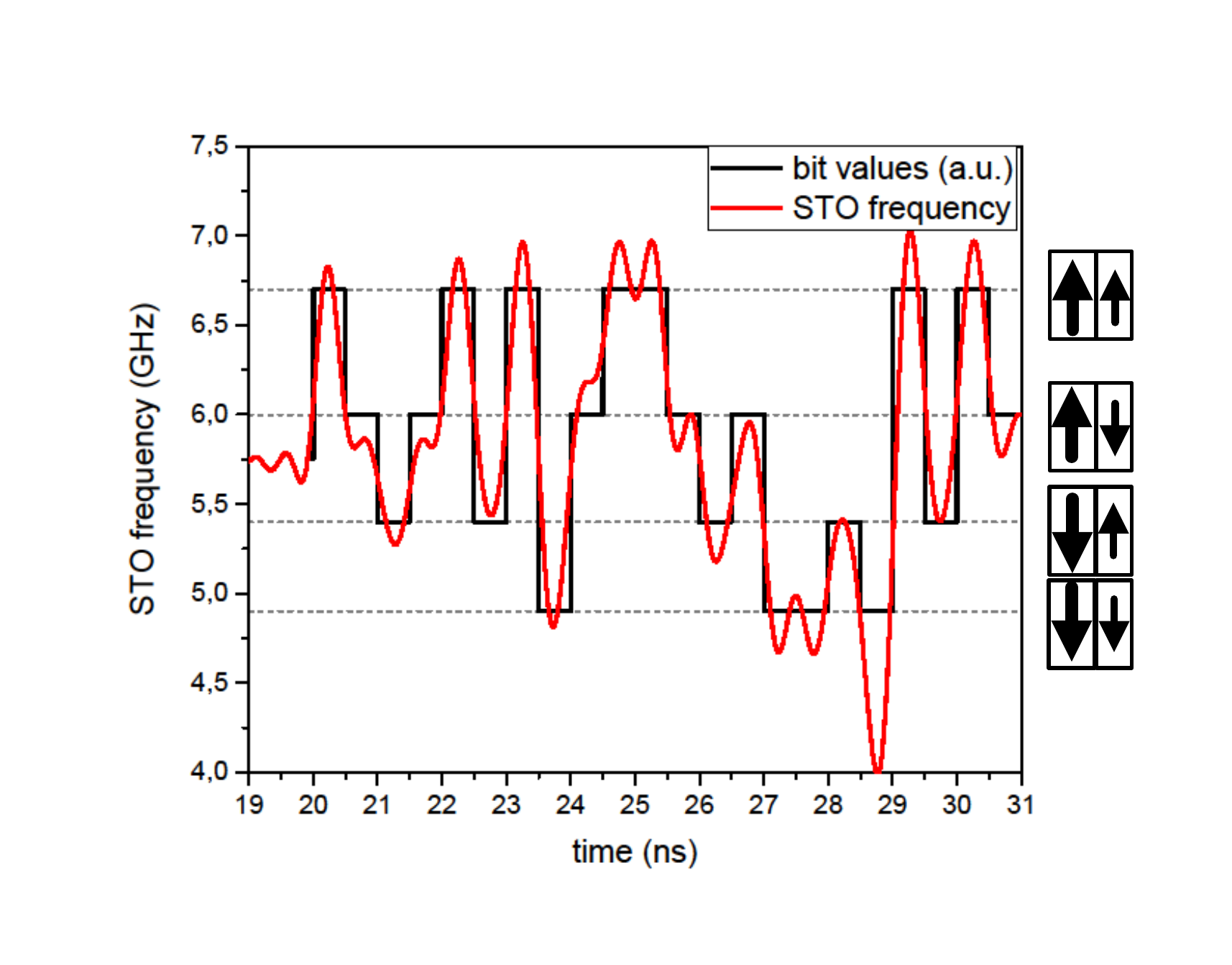}
\caption{Simulation of storage media readout. The red line represents instantaneous STO frequency as a function of time. The black line represents approximate expected frequencies based on the average magnetic field from the two-bit state.}
\label{fig1}
\end{figure}

After analyzing the system behavior in stationary states, we performed a simulation of a dynamic hard disk media readout. An example sequence of several two-bit pairs was generated and the corresponding magnetic fields were applied to the left and right parts of the free layer. The field amplitude remained the same as in the previous subsection and the state was changed every 0.5 ns, which corresponds to a nominal data transfer rate of 2 Gb/s. We note that the actual data transfer rate is higher because of the fact that two bit tracks are read simultaneously and equals 4 Gb/s in this scenario. The setup we propose can thus be understood as a method to enhance data transfer rate value while keeping the physical speed of the device the same. 

Figure 4 presents an instantaneous frequency of the simulated STO device (red line) as a function of time, with bit pair values (black line) denoted symbolically as approximate frequency levels corresponding to their respective field amplitudes. On the right side of the figure, a schematic picture of each two-bit scenario can be seen. The instantaneous frequency was defined as a derivative of instantaneous STO phase with respect to time and it was calculated using a Hilbert transform of the system resistance and a 2 GHz low-pass filter \cite{sato2012simulations,checinski2017spin}. One can see that the device frequency closely follows the expected levels and allows to obtain both bit values easily, leading to a successful readout characterized by a data transfer ratio twice as high as in the case of a single-track readout.

\section{Influence of magnetic noise}

The micromagnetic simulations described so far have been performed under the assumption of zero temperature. However, one of the main advantages of the proposed setup is, supposedly, its increased resilience to magnetic noise which is an effect inherently connected with the presence of non-zero temperature. Therefore, it would be useful to perform additional simulations including also thermal effects, to check whether the assumed magnetic field inhomogeneity can result in additional line broadening in the FFT spectrum of the device response. Additionally, the SNR ratio of an STO reader depends on the difference between readout frequency levels \cite{braganca2010nanoscale} and the setup we propose introduces additional levels inside the readout range. As a result, an estimation of the respective SNR levels is needed in order to compare the influence of this effect and volume decrease effect.

\subsection{Micromagnetic simulation of thermal effects in a two-bit reader} 

In order to investigate the linewidth changes in the two-bit reader compared to the single-bit reader scenario, we performed micromagnetic simulations of the device described in section II in the presence of thermal effects. Since the device linewidth can depend on the current amplitude, a set of simulations for different current amplitudes was carried out. The positive magnetic field of +1.3 mT was chosen, and it was applied to the free layer of the device either in a uniform manner, or as the "10" scenario where field in the left part of the device was equal to +5 mT and field in the right part of the device was equal to -5 mT. The thermal effects were modelled as a white Gaussian noise which was added to each magnetization cell vector independently in each simulation step. After initial magnetization relaxation, the system was simulated for 50 nanoseconds and the FFT response was calculated, similarly to what was presented in section II A. We repeated the simulation for each current amplitude four times using different random number generator seeds and the results were averaged to better capture the response linewidth. The exact linewidth value was obtained from fitting a Gaussian curve to the averaged result. 

Figure 5. presents the simulated full width at half maximum (FWHM) of the device response spectrum as a function of current amplitude, for both the uniform case (black line and points) and the non-uniform "10" case (red line and points). One can see that the linewidth values in both cases remain similar for a wide range of current amplitudes. In fact, the linewidth in the double-field scenario turns out to be about 30$\%$ lower on average. We hypothesize that the exact physical origin of this phenomenon may be connected with the free layer becoming more magnetically stiff or with the overall precession cone angle decreasing in the double field scenario. Nonetheless, we conclude that the typical linewidth of the STO reader in the setup presented in Figure 1. should be at most comparable to that of a single-bit reader, and that no significant line broadening has occurred despite the external field inhomogeneity. We also note that the curves presented in Fig. 5. show qualitative agreement with the expected dependence for in-plane magnetized STOs \cite{slavin2009nonlinear}.
 
 \begin{figure}
\centering
\includegraphics[width=8.8cm]{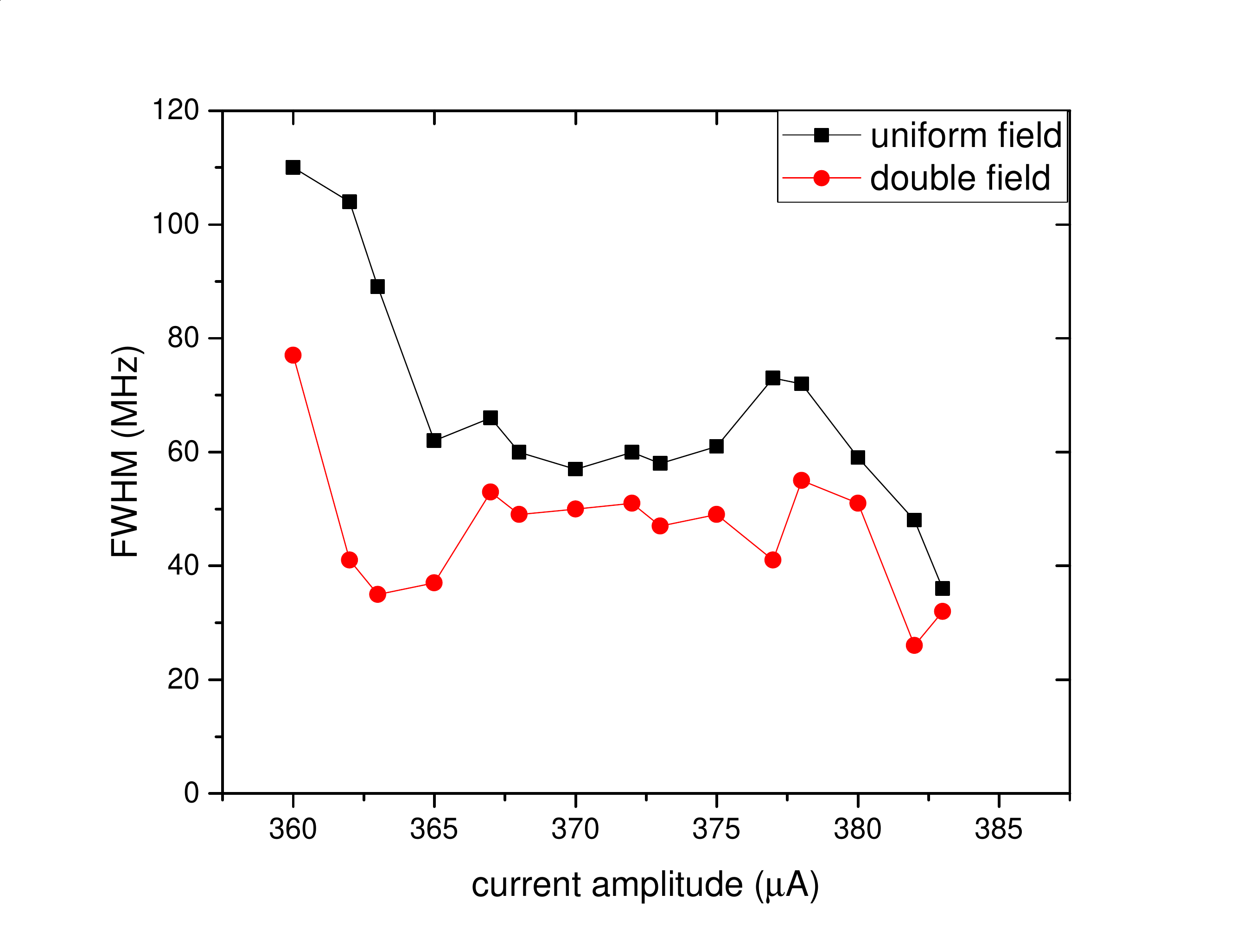}
\caption{Simulated FWHM of the device response spectrum as a function of spin-polarized current amplitude for uniform field (black) and double field (red) scenarios.}
\label{fig1}
\end{figure}

\subsection{Signal to noise ratio estimations}

The SNR ratio for an STO-based magnetic recording reader can be expressed as \cite{braganca2010nanoscale}:
\begin{equation}
SNR = const \cdot ln\left(2ln(2)\left(\frac{\Delta f}{\tau}\right)^2\right),
\end{equation}
where $\Delta f$ refers to the difference between frequency levels used for readout and $\tau$ is the FWHM of the magnetization response spectrum. In the usually assumed setup, only two frequency levels are utilized, corresponding to either positive or negative external field value. On the contrary, in the setup we propose there are four possible frequency levels. Therefore, the minimal difference between readout frequency levels in our setup $\Delta f'$ is lower than the corresponding $\Delta f$ of the same device. The exact value of this decrease depends on the system parameters. Here, we will use an approximation $\Delta f' = \Delta f / 3$, which is equivalent to assuming that the additional frequency levels visible in Figs. 2 and 4 are evenly spread. At the same time, the linewidth of the device has been shown to be inversely proportional to the free layer volume $V$ \cite{mizushima2010signal}, which can be expressed as $\tau = \beta_0 / V$, where $\beta_0$ is a certain volume-independent constant. As a result, we can obtain the following condition for equal SNR values in uniform and double field scenarios:
\begin{equation}
r' = \sqrt{\frac{3\beta'_0}{\beta_0}}r,
\end{equation}
where $r$ and $r'$ refer to the lateral sizes of the (circular) device in uniform and double field scenarios while $\beta_0$ and $\beta'_0$ refer to the respective size-independent constants determining FWHM levels. If we assume that $\beta_0 = \beta'_0$ (size-independent factors the same in both scenarios), a double-bit reader of 50 nm diameter is expected to have the same SNR as a single-bit reader of approximately 29 nm diameter, while retaining twice as high data transfer rate. Moving further, if we assume $\beta'_0$ to be 30$\%$ lower than $\beta_0$, as indicated by the simulations described in the previous paragraph, a double-bit reader of 50 nm diameter will have the same SNR as a single-bit reader of approximately 35 nm diameter, which means that in a $w_1$ = 30 nm, $w_2$ = 20 nm usage scenario both SNR and data transfer rate would be improved. These calculations do not take into account other effects of decreased size, such as different current threshold, decreased dipolar coupling field between the free layer and other junction layers, or changes in shape anisotropy. Nevertheless, the obtained results show that the proposed setup can lead to substantial benefits from the STO usage in HDD readout point of view. Depending on particular technological requirements and choice of STO size, these benefits could mean increased data transfer rate, increased SNR value or a combination of both of these. 
\\

\section{Summary}

We have presented a novel setup for simultaneous detection of two storage bits with spin-torque oscillator. The micromagnetic simulations have shown that the magnetization response can remain unimodal under non-uniform external magnetic field and that differentiation between "01" and  "10" is possible for asymmetric geometry of the setup. We have performed a successful readout simulation with all four combinations of bit values included. We have calculated the oscillator linewidth as well as estimated the signal to noise ratios in the double field scenario and found that the proposed setup can prove beneficial in terms of higher data transfer rate or increased resilience to thermal noise. 

\section*{Acknowledgements}
We acknowledge the grant Preludium 2015/17/N/ST7/ 03749 by National Science Center, Poland. J.Ch. acknowledges the scholarship under Marian Smoluchowski Krakow Research Consortium KNOW programme. Numerical calculations were supported by PL-GRID infrastructure.



\bibliographystyle{apsrev4-1}
\bibliography{bibliography}

\begin{thebibliography}{32}%
\makeatletter
\providecommand \@ifxundefined [1]{%
 \@ifx{#1\undefined}
}%
\providecommand \@ifnum [1]{%
 \ifnum #1\expandafter \@firstoftwo
 \else \expandafter \@secondoftwo
 \fi
}%
\providecommand \@ifx [1]{%
 \ifx #1\expandafter \@firstoftwo
 \else \expandafter \@secondoftwo
 \fi
}%
\providecommand \natexlab [1]{#1}%
\providecommand \enquote  [1]{``#1''}%
\providecommand \bibnamefont  [1]{#1}%
\providecommand \bibfnamefont [1]{#1}%
\providecommand \citenamefont [1]{#1}%
\providecommand \href@noop [0]{\@secondoftwo}%
\providecommand \href [0]{\begingroup \@sanitize@url \@href}%
\providecommand \@href[1]{\@@startlink{#1}\@@href}%
\providecommand \@@href[1]{\endgroup#1\@@endlink}%
\providecommand \@sanitize@url [0]{\catcode `\\12\catcode `\$12\catcode
  `\&12\catcode `\#12\catcode `\^12\catcode `\_12\catcode `\%12\relax}%
\providecommand \@@startlink[1]{}%
\providecommand \@@endlink[0]{}%
\providecommand \url  [0]{\begingroup\@sanitize@url \@url }%
\providecommand \@url [1]{\endgroup\@href {#1}{\urlprefix }}%
\providecommand \urlprefix  [0]{URL }%
\providecommand \Eprint [0]{\href }%
\providecommand \doibase [0]{http://dx.doi.org/}%
\providecommand \selectlanguage [0]{\@gobble}%
\providecommand \bibinfo  [0]{\@secondoftwo}%
\providecommand \bibfield  [0]{\@secondoftwo}%
\providecommand \translation [1]{[#1]}%
\providecommand \BibitemOpen [0]{}%
\providecommand \bibitemStop [0]{}%
\providecommand \bibitemNoStop [0]{.\EOS\space}%
\providecommand \EOS [0]{\spacefactor3000\relax}%
\providecommand \BibitemShut  [1]{\csname bibitem#1\endcsname}%
\let\auto@bib@innerbib\@empty
\bibitem [{\citenamefont {Tsoi}\ \emph {et~al.}(1998)\citenamefont {Tsoi},
  \citenamefont {Jansen}, \citenamefont {Bass}, \citenamefont {Chiang},
  \citenamefont {Seck}, \citenamefont {Tsoi},\ and\ \citenamefont
  {Wyder}}]{tsoi1998excitation}%
  \BibitemOpen
  \bibfield  {author} {\bibinfo {author} {\bibfnamefont {M.}~\bibnamefont
  {Tsoi}}, \bibinfo {author} {\bibfnamefont {A.}~\bibnamefont {Jansen}},
  \bibinfo {author} {\bibfnamefont {J.}~\bibnamefont {Bass}}, \bibinfo {author}
  {\bibfnamefont {W.-C.}\ \bibnamefont {Chiang}}, \bibinfo {author}
  {\bibfnamefont {M.}~\bibnamefont {Seck}}, \bibinfo {author} {\bibfnamefont
  {V.}~\bibnamefont {Tsoi}}, \ and\ \bibinfo {author} {\bibfnamefont
  {P.}~\bibnamefont {Wyder}},\ }\href@noop {} {\bibfield  {journal} {\bibinfo
  {journal} {Physical Review Letters}\ }\textbf {\bibinfo {volume} {80}},\
  \bibinfo {pages} {4281} (\bibinfo {year} {1998})}\BibitemShut {NoStop}%
\bibitem [{\citenamefont {Rippard}\ \emph {et~al.}(2005)\citenamefont
  {Rippard}, \citenamefont {Pufall}, \citenamefont {Kaka}, \citenamefont
  {Silva}, \citenamefont {Russek},\ and\ \citenamefont
  {Katine}}]{rippard2005injection}%
  \BibitemOpen
  \bibfield  {author} {\bibinfo {author} {\bibfnamefont {W.}~\bibnamefont
  {Rippard}}, \bibinfo {author} {\bibfnamefont {M.}~\bibnamefont {Pufall}},
  \bibinfo {author} {\bibfnamefont {S.}~\bibnamefont {Kaka}}, \bibinfo {author}
  {\bibfnamefont {T.}~\bibnamefont {Silva}}, \bibinfo {author} {\bibfnamefont
  {S.}~\bibnamefont {Russek}}, \ and\ \bibinfo {author} {\bibfnamefont
  {J.}~\bibnamefont {Katine}},\ }\href@noop {} {\bibfield  {journal} {\bibinfo
  {journal} {Physical Review Letters}\ }\textbf {\bibinfo {volume} {95}},\
  \bibinfo {pages} {067203} (\bibinfo {year} {2005})}\BibitemShut {NoStop}%
\bibitem [{\citenamefont {Kiselev}\ \emph {et~al.}(2003)\citenamefont
  {Kiselev}, \citenamefont {Sankey}, \citenamefont {Krivorotov}, \citenamefont
  {Emley}, \citenamefont {Schoelkopf}, \citenamefont {Buhrman},\ and\
  \citenamefont {Ralph}}]{kiselev2003microwave}%
  \BibitemOpen
  \bibfield  {author} {\bibinfo {author} {\bibfnamefont {S.~I.}\ \bibnamefont
  {Kiselev}}, \bibinfo {author} {\bibfnamefont {J.}~\bibnamefont {Sankey}},
  \bibinfo {author} {\bibfnamefont {I.}~\bibnamefont {Krivorotov}}, \bibinfo
  {author} {\bibfnamefont {N.}~\bibnamefont {Emley}}, \bibinfo {author}
  {\bibfnamefont {R.}~\bibnamefont {Schoelkopf}}, \bibinfo {author}
  {\bibfnamefont {R.}~\bibnamefont {Buhrman}}, \ and\ \bibinfo {author}
  {\bibfnamefont {D.}~\bibnamefont {Ralph}},\ }\href@noop {} {\bibfield
  {journal} {\bibinfo  {journal} {Nature}\ }\textbf {\bibinfo {volume} {425}},\
  \bibinfo {pages} {380} (\bibinfo {year} {2003})}\BibitemShut {NoStop}%
\bibitem [{\citenamefont {Houssameddine}\ \emph {et~al.}(2007)\citenamefont
  {Houssameddine}, \citenamefont {Ebels}, \citenamefont {Dela{\"e}t},
  \citenamefont {Rodmacq}, \citenamefont {Firastrau}, \citenamefont
  {Ponthenier}, \citenamefont {Brunet}, \citenamefont {Thirion}, \citenamefont
  {Michel}, \citenamefont {Prejbeanu-Buda} \emph
  {et~al.}}]{houssameddine2007spin}%
  \BibitemOpen
  \bibfield  {author} {\bibinfo {author} {\bibfnamefont {D.}~\bibnamefont
  {Houssameddine}}, \bibinfo {author} {\bibfnamefont {U.}~\bibnamefont
  {Ebels}}, \bibinfo {author} {\bibfnamefont {B.}~\bibnamefont {Dela{\"e}t}},
  \bibinfo {author} {\bibfnamefont {B.}~\bibnamefont {Rodmacq}}, \bibinfo
  {author} {\bibfnamefont {I.}~\bibnamefont {Firastrau}}, \bibinfo {author}
  {\bibfnamefont {F.}~\bibnamefont {Ponthenier}}, \bibinfo {author}
  {\bibfnamefont {M.}~\bibnamefont {Brunet}}, \bibinfo {author} {\bibfnamefont
  {C.}~\bibnamefont {Thirion}}, \bibinfo {author} {\bibfnamefont {J.-P.}\
  \bibnamefont {Michel}}, \bibinfo {author} {\bibfnamefont {L.}~\bibnamefont
  {Prejbeanu-Buda}},  \emph {et~al.},\ }\href@noop {} {\bibfield  {journal}
  {\bibinfo  {journal} {Nature Materials}\ }\textbf {\bibinfo {volume} {6}},\
  \bibinfo {pages} {447} (\bibinfo {year} {2007})}\BibitemShut {NoStop}%
\bibitem [{\citenamefont {Deac}\ \emph {et~al.}(2008)\citenamefont {Deac},
  \citenamefont {Fukushima}, \citenamefont {Kubota}, \citenamefont {Maehara},
  \citenamefont {Suzuki}, \citenamefont {Yuasa}, \citenamefont {Nagamine},
  \citenamefont {Tsunekawa}, \citenamefont {Djayaprawira},\ and\ \citenamefont
  {Watanabe}}]{deac2008bias}%
  \BibitemOpen
  \bibfield  {author} {\bibinfo {author} {\bibfnamefont {A.~M.}\ \bibnamefont
  {Deac}}, \bibinfo {author} {\bibfnamefont {A.}~\bibnamefont {Fukushima}},
  \bibinfo {author} {\bibfnamefont {H.}~\bibnamefont {Kubota}}, \bibinfo
  {author} {\bibfnamefont {H.}~\bibnamefont {Maehara}}, \bibinfo {author}
  {\bibfnamefont {Y.}~\bibnamefont {Suzuki}}, \bibinfo {author} {\bibfnamefont
  {S.}~\bibnamefont {Yuasa}}, \bibinfo {author} {\bibfnamefont
  {Y.}~\bibnamefont {Nagamine}}, \bibinfo {author} {\bibfnamefont
  {K.}~\bibnamefont {Tsunekawa}}, \bibinfo {author} {\bibfnamefont {D.~D.}\
  \bibnamefont {Djayaprawira}}, \ and\ \bibinfo {author} {\bibfnamefont
  {N.}~\bibnamefont {Watanabe}},\ }\href@noop {} {\bibfield  {journal}
  {\bibinfo  {journal} {Nature Physics}\ }\textbf {\bibinfo {volume} {4}},\
  \bibinfo {pages} {803} (\bibinfo {year} {2008})}\BibitemShut {NoStop}%
\bibitem [{\citenamefont {Krivorotov}\ \emph {et~al.}(2007)\citenamefont
  {Krivorotov}, \citenamefont {Berkov}, \citenamefont {Gorn}, \citenamefont
  {Emley}, \citenamefont {Sankey}, \citenamefont {Ralph},\ and\ \citenamefont
  {Buhrman}}]{krivorotov2007large}%
  \BibitemOpen
  \bibfield  {author} {\bibinfo {author} {\bibfnamefont {I.}~\bibnamefont
  {Krivorotov}}, \bibinfo {author} {\bibfnamefont {D.}~\bibnamefont {Berkov}},
  \bibinfo {author} {\bibfnamefont {N.}~\bibnamefont {Gorn}}, \bibinfo {author}
  {\bibfnamefont {N.}~\bibnamefont {Emley}}, \bibinfo {author} {\bibfnamefont
  {J.}~\bibnamefont {Sankey}}, \bibinfo {author} {\bibfnamefont
  {D.}~\bibnamefont {Ralph}}, \ and\ \bibinfo {author} {\bibfnamefont
  {R.}~\bibnamefont {Buhrman}},\ }\href@noop {} {\bibfield  {journal} {\bibinfo
   {journal} {Physical Review B}\ }\textbf {\bibinfo {volume} {76}},\ \bibinfo
  {pages} {024418} (\bibinfo {year} {2007})}\BibitemShut {NoStop}%
\bibitem [{\citenamefont {Kim}(2012)}]{kima2012spin}%
  \BibitemOpen
  \bibfield  {author} {\bibinfo {author} {\bibfnamefont {J.-V.}\ \bibnamefont
  {Kim}},\ }\href@noop {} {\bibfield  {journal} {\bibinfo  {journal} {Solid
  State Physics}\ }\textbf {\bibinfo {volume} {63}},\ \bibinfo {pages} {217}
  (\bibinfo {year} {2012})}\BibitemShut {NoStop}%
\bibitem [{\citenamefont {Dumas}\ \emph {et~al.}(2014)\citenamefont {Dumas},
  \citenamefont {Sani}, \citenamefont {Mohseni}, \citenamefont {Iacocca},
  \citenamefont {Pogoryelov}, \citenamefont {Muduli}, \citenamefont {Chung},
  \citenamefont {Durrenfeld},\ and\ \citenamefont {Akerman}}]{dumas2014recent}%
  \BibitemOpen
  \bibfield  {author} {\bibinfo {author} {\bibfnamefont {R.~K.}\ \bibnamefont
  {Dumas}}, \bibinfo {author} {\bibfnamefont {S.~R.}\ \bibnamefont {Sani}},
  \bibinfo {author} {\bibfnamefont {S.~M.}\ \bibnamefont {Mohseni}}, \bibinfo
  {author} {\bibfnamefont {E.}~\bibnamefont {Iacocca}}, \bibinfo {author}
  {\bibfnamefont {Y.}~\bibnamefont {Pogoryelov}}, \bibinfo {author}
  {\bibfnamefont {P.~K.}\ \bibnamefont {Muduli}}, \bibinfo {author}
  {\bibfnamefont {S.}~\bibnamefont {Chung}}, \bibinfo {author} {\bibfnamefont
  {P.}~\bibnamefont {Durrenfeld}}, \ and\ \bibinfo {author} {\bibfnamefont
  {J.}~\bibnamefont {Akerman}},\ }\href@noop {} {\bibfield  {journal} {\bibinfo
   {journal} {IEEE transactions on magnetics}\ }\textbf {\bibinfo {volume}
  {50}},\ \bibinfo {pages} {1} (\bibinfo {year} {2014})}\BibitemShut {NoStop}%
\bibitem [{\citenamefont {Skowro{\'n}ski}\ \emph {et~al.}(2012)\citenamefont
  {Skowro{\'n}ski}, \citenamefont {Stobiecki}, \citenamefont {Wrona},
  \citenamefont {Reiss},\ and\ \citenamefont {van
  Dijken}}]{skowronski2012zero}%
  \BibitemOpen
  \bibfield  {author} {\bibinfo {author} {\bibfnamefont {W.}~\bibnamefont
  {Skowro{\'n}ski}}, \bibinfo {author} {\bibfnamefont {T.}~\bibnamefont
  {Stobiecki}}, \bibinfo {author} {\bibfnamefont {J.}~\bibnamefont {Wrona}},
  \bibinfo {author} {\bibfnamefont {G.}~\bibnamefont {Reiss}}, \ and\ \bibinfo
  {author} {\bibfnamefont {S.}~\bibnamefont {van Dijken}},\ }\href@noop {}
  {\bibfield  {journal} {\bibinfo  {journal} {Applied Physics Express}\
  }\textbf {\bibinfo {volume} {5}},\ \bibinfo {pages} {063005} (\bibinfo {year}
  {2012})}\BibitemShut {NoStop}%
\bibitem [{\citenamefont {Zeng}\ \emph {et~al.}(2013)\citenamefont {Zeng},
  \citenamefont {Finocchio}, \citenamefont {Zhang}, \citenamefont {Amiri},
  \citenamefont {Katine}, \citenamefont {Krivorotov}, \citenamefont {Huai},
  \citenamefont {Langer}, \citenamefont {Azzerboni}, \citenamefont {Wang} \emph
  {et~al.}}]{zeng2013ultralow}%
  \BibitemOpen
  \bibfield  {author} {\bibinfo {author} {\bibfnamefont {Z.}~\bibnamefont
  {Zeng}}, \bibinfo {author} {\bibfnamefont {G.}~\bibnamefont {Finocchio}},
  \bibinfo {author} {\bibfnamefont {B.}~\bibnamefont {Zhang}}, \bibinfo
  {author} {\bibfnamefont {P.~K.}\ \bibnamefont {Amiri}}, \bibinfo {author}
  {\bibfnamefont {J.~A.}\ \bibnamefont {Katine}}, \bibinfo {author}
  {\bibfnamefont {I.~N.}\ \bibnamefont {Krivorotov}}, \bibinfo {author}
  {\bibfnamefont {Y.}~\bibnamefont {Huai}}, \bibinfo {author} {\bibfnamefont
  {J.}~\bibnamefont {Langer}}, \bibinfo {author} {\bibfnamefont
  {B.}~\bibnamefont {Azzerboni}}, \bibinfo {author} {\bibfnamefont {K.~L.}\
  \bibnamefont {Wang}},  \emph {et~al.},\ }\href@noop {} {\bibfield  {journal}
  {\bibinfo  {journal} {Scientific Reports}\ }\textbf {\bibinfo {volume} {3}}
  (\bibinfo {year} {2013})}\BibitemShut {NoStop}%
\bibitem [{\citenamefont {Madami}\ \emph {et~al.}(2011)\citenamefont {Madami},
  \citenamefont {Bonetti}, \citenamefont {Consolo}, \citenamefont {Tacchi},
  \citenamefont {Carlotti}, \citenamefont {Gubbiotti}, \citenamefont {Mancoff},
  \citenamefont {Yar},\ and\ \citenamefont {{\AA}kerman}}]{madami2011direct}%
  \BibitemOpen
  \bibfield  {author} {\bibinfo {author} {\bibfnamefont {M.}~\bibnamefont
  {Madami}}, \bibinfo {author} {\bibfnamefont {S.}~\bibnamefont {Bonetti}},
  \bibinfo {author} {\bibfnamefont {G.}~\bibnamefont {Consolo}}, \bibinfo
  {author} {\bibfnamefont {S.}~\bibnamefont {Tacchi}}, \bibinfo {author}
  {\bibfnamefont {G.}~\bibnamefont {Carlotti}}, \bibinfo {author}
  {\bibfnamefont {G.}~\bibnamefont {Gubbiotti}}, \bibinfo {author}
  {\bibfnamefont {F.}~\bibnamefont {Mancoff}}, \bibinfo {author} {\bibfnamefont
  {M.~A.}\ \bibnamefont {Yar}}, \ and\ \bibinfo {author} {\bibfnamefont
  {J.}~\bibnamefont {{\AA}kerman}},\ }\href@noop {} {\bibfield  {journal}
  {\bibinfo  {journal} {Nature Nanotechnology}\ }\textbf {\bibinfo {volume}
  {6}},\ \bibinfo {pages} {635} (\bibinfo {year} {2011})}\BibitemShut {NoStop}%
\bibitem [{\citenamefont {Demidov}\ \emph {et~al.}(2010)\citenamefont
  {Demidov}, \citenamefont {Urazhdin},\ and\ \citenamefont
  {Demokritov}}]{demidov2010direct}%
  \BibitemOpen
  \bibfield  {author} {\bibinfo {author} {\bibfnamefont {V.~E.}\ \bibnamefont
  {Demidov}}, \bibinfo {author} {\bibfnamefont {S.}~\bibnamefont {Urazhdin}}, \
  and\ \bibinfo {author} {\bibfnamefont {S.~O.}\ \bibnamefont {Demokritov}},\
  }\href@noop {} {\bibfield  {journal} {\bibinfo  {journal} {Nature Materials}\
  }\textbf {\bibinfo {volume} {9}},\ \bibinfo {pages} {984} (\bibinfo {year}
  {2010})}\BibitemShut {NoStop}%
\bibitem [{\citenamefont {Kudo}\ \emph {et~al.}(2009)\citenamefont {Kudo},
  \citenamefont {Nagasawa}, \citenamefont {Sato},\ and\ \citenamefont
  {Mizushima}}]{kudo2009measurement}%
  \BibitemOpen
  \bibfield  {author} {\bibinfo {author} {\bibfnamefont {K.}~\bibnamefont
  {Kudo}}, \bibinfo {author} {\bibfnamefont {T.}~\bibnamefont {Nagasawa}},
  \bibinfo {author} {\bibfnamefont {R.}~\bibnamefont {Sato}}, \ and\ \bibinfo
  {author} {\bibfnamefont {K.}~\bibnamefont {Mizushima}},\ }\href@noop {}
  {\bibfield  {journal} {\bibinfo  {journal} {Applied Physics Letters}\
  }\textbf {\bibinfo {volume} {95}},\ \bibinfo {pages} {022507} (\bibinfo
  {year} {2009})}\BibitemShut {NoStop}%
\bibitem [{\citenamefont {Braganca}\ \emph {et~al.}(2010)\citenamefont
  {Braganca}, \citenamefont {Gurney}, \citenamefont {Wilson}, \citenamefont
  {Katine}, \citenamefont {Maat},\ and\ \citenamefont
  {Childress}}]{braganca2010nanoscale}%
  \BibitemOpen
  \bibfield  {author} {\bibinfo {author} {\bibfnamefont {P.}~\bibnamefont
  {Braganca}}, \bibinfo {author} {\bibfnamefont {B.}~\bibnamefont {Gurney}},
  \bibinfo {author} {\bibfnamefont {B.}~\bibnamefont {Wilson}}, \bibinfo
  {author} {\bibfnamefont {J.}~\bibnamefont {Katine}}, \bibinfo {author}
  {\bibfnamefont {S.}~\bibnamefont {Maat}}, \ and\ \bibinfo {author}
  {\bibfnamefont {J.}~\bibnamefont {Childress}},\ }\href@noop {} {\bibfield
  {journal} {\bibinfo  {journal} {Nanotechnology}\ }\textbf {\bibinfo {volume}
  {21}},\ \bibinfo {pages} {235202} (\bibinfo {year} {2010})}\BibitemShut
  {NoStop}%
\bibitem [{\citenamefont {Sato}\ \emph {et~al.}(2012)\citenamefont {Sato},
  \citenamefont {Kudo}, \citenamefont {Nagasawa}, \citenamefont {Suto},\ and\
  \citenamefont {Mizushima}}]{sato2012simulations}%
  \BibitemOpen
  \bibfield  {author} {\bibinfo {author} {\bibfnamefont {R.}~\bibnamefont
  {Sato}}, \bibinfo {author} {\bibfnamefont {K.}~\bibnamefont {Kudo}}, \bibinfo
  {author} {\bibfnamefont {T.}~\bibnamefont {Nagasawa}}, \bibinfo {author}
  {\bibfnamefont {H.}~\bibnamefont {Suto}}, \ and\ \bibinfo {author}
  {\bibfnamefont {K.}~\bibnamefont {Mizushima}},\ }\href@noop {} {\bibfield
  {journal} {\bibinfo  {journal} {IEEE Transactions on Magnetics}\ }\textbf
  {\bibinfo {volume} {48}},\ \bibinfo {pages} {1758} (\bibinfo {year}
  {2012})}\BibitemShut {NoStop}%
\bibitem [{\citenamefont {Suto}\ \emph {et~al.}(2014)\citenamefont {Suto},
  \citenamefont {Nagasawa}, \citenamefont {Kudo}, \citenamefont {Mizushima},\
  and\ \citenamefont {Sato}}]{suto2014nanoscale}%
  \BibitemOpen
  \bibfield  {author} {\bibinfo {author} {\bibfnamefont {H.}~\bibnamefont
  {Suto}}, \bibinfo {author} {\bibfnamefont {T.}~\bibnamefont {Nagasawa}},
  \bibinfo {author} {\bibfnamefont {K.}~\bibnamefont {Kudo}}, \bibinfo {author}
  {\bibfnamefont {K.}~\bibnamefont {Mizushima}}, \ and\ \bibinfo {author}
  {\bibfnamefont {R.}~\bibnamefont {Sato}},\ }\href@noop {} {\bibfield
  {journal} {\bibinfo  {journal} {Nanotechnology}\ }\textbf {\bibinfo {volume}
  {25}},\ \bibinfo {pages} {245501} (\bibinfo {year} {2014})}\BibitemShut
  {NoStop}%
\bibitem [{\citenamefont {Checinski}\ \emph {et~al.}(2017)\citenamefont
  {Checinski}, \citenamefont {Frankowski},\ and\ \citenamefont
  {Stobiecki}}]{checinski2017spin}%
  \BibitemOpen
  \bibfield  {author} {\bibinfo {author} {\bibfnamefont {J.}~\bibnamefont
  {Checinski}}, \bibinfo {author} {\bibfnamefont {M.}~\bibnamefont
  {Frankowski}}, \ and\ \bibinfo {author} {\bibfnamefont {T.}~\bibnamefont
  {Stobiecki}},\ }\href@noop {} {\bibfield  {journal} {\bibinfo  {journal}
  {IEEE Transactions on Magnetics}\ }\textbf {\bibinfo {volume} {53}} (\bibinfo
  {year} {2017})}\BibitemShut {NoStop}%
\bibitem [{\citenamefont {Mizushima.}\ \emph {et~al.}(2010)\citenamefont
  {Mizushima.}, \citenamefont {Kudo}, \citenamefont {Nagasawa},\ and\
  \citenamefont {Sato}}]{mizushima2010signal}%
  \BibitemOpen
  \bibfield  {author} {\bibinfo {author} {\bibfnamefont {K.}~\bibnamefont
  {Mizushima.}}, \bibinfo {author} {\bibfnamefont {K.}~\bibnamefont {Kudo}},
  \bibinfo {author} {\bibfnamefont {T.}~\bibnamefont {Nagasawa}}, \ and\
  \bibinfo {author} {\bibfnamefont {R.}~\bibnamefont {Sato}},\ }\href@noop {}
  {\bibfield  {journal} {\bibinfo  {journal} {Journal of Applied Physics}\
  }\textbf {\bibinfo {volume} {107}},\ \bibinfo {pages} {063904} (\bibinfo
  {year} {2010})}\BibitemShut {NoStop}%
\bibitem [{\citenamefont {Kanao}\ \emph {et~al.}(2016)\citenamefont {Kanao},
  \citenamefont {Nagasawa}, \citenamefont {Kudo}, \citenamefont {Suto},
  \citenamefont {Yamagishi}, \citenamefont {Mizushima},\ and\ \citenamefont
  {Sato}}]{kanao2016effects}%
  \BibitemOpen
  \bibfield  {author} {\bibinfo {author} {\bibfnamefont {T.}~\bibnamefont
  {Kanao}}, \bibinfo {author} {\bibfnamefont {T.}~\bibnamefont {Nagasawa}},
  \bibinfo {author} {\bibfnamefont {K.}~\bibnamefont {Kudo}}, \bibinfo {author}
  {\bibfnamefont {H.}~\bibnamefont {Suto}}, \bibinfo {author} {\bibfnamefont
  {M.}~\bibnamefont {Yamagishi}}, \bibinfo {author} {\bibfnamefont
  {K.}~\bibnamefont {Mizushima}}, \ and\ \bibinfo {author} {\bibfnamefont
  {R.}~\bibnamefont {Sato}},\ }\href@noop {} {\bibfield  {journal} {\bibinfo
  {journal} {Applied Physics Express}\ }\textbf {\bibinfo {volume} {9}},\
  \bibinfo {pages} {113001} (\bibinfo {year} {2016})}\BibitemShut {NoStop}%
\bibitem [{\citenamefont {Nagasawa}\ \emph {et~al.}(2016)\citenamefont
  {Nagasawa}, \citenamefont {Suto}, \citenamefont {Kudo}, \citenamefont
  {Mizushima},\ and\ \citenamefont {Sato}}]{nagasawa2016response}%
  \BibitemOpen
  \bibfield  {author} {\bibinfo {author} {\bibfnamefont {T.}~\bibnamefont
  {Nagasawa}}, \bibinfo {author} {\bibfnamefont {H.}~\bibnamefont {Suto}},
  \bibinfo {author} {\bibfnamefont {K.}~\bibnamefont {Kudo}}, \bibinfo {author}
  {\bibfnamefont {K.}~\bibnamefont {Mizushima}}, \ and\ \bibinfo {author}
  {\bibfnamefont {R.}~\bibnamefont {Sato}},\ }\href@noop {} {\bibfield
  {journal} {\bibinfo  {journal} {Applied Physics Express}\ }\textbf {\bibinfo
  {volume} {9}},\ \bibinfo {pages} {113002} (\bibinfo {year}
  {2016})}\BibitemShut {NoStop}%
\bibitem [{\citenamefont {Chao}\ \emph {et~al.}(2017)\citenamefont {Chao},
  \citenamefont {Jamali},\ and\ \citenamefont {Wang}}]{chao2017scaling}%
  \BibitemOpen
  \bibfield  {author} {\bibinfo {author} {\bibfnamefont {X.}~\bibnamefont
  {Chao}}, \bibinfo {author} {\bibfnamefont {M.}~\bibnamefont {Jamali}}, \ and\
  \bibinfo {author} {\bibfnamefont {J.-P.}\ \bibnamefont {Wang}},\ }\href@noop
  {} {\bibfield  {journal} {\bibinfo  {journal} {AIP Advances}\ }\textbf
  {\bibinfo {volume} {7}},\ \bibinfo {pages} {056624} (\bibinfo {year}
  {2017})}\BibitemShut {NoStop}%
\bibitem [{\citenamefont {Li}\ and\ \citenamefont
  {Wei}(2016)}]{li2016micromagnetic}%
  \BibitemOpen
  \bibfield  {author} {\bibinfo {author} {\bibfnamefont {J.}~\bibnamefont
  {Li}}\ and\ \bibinfo {author} {\bibfnamefont {D.}~\bibnamefont {Wei}},\
  }\href@noop {} {\bibfield  {journal} {\bibinfo  {journal} {IEEE Magnetics
  Letters}\ }\textbf {\bibinfo {volume} {7}},\ \bibinfo {pages} {1} (\bibinfo
  {year} {2016})}\BibitemShut {NoStop}%
\bibitem [{\citenamefont {Xia}\ \emph {et~al.}(2017)\citenamefont {Xia},
  \citenamefont {Zheng}, \citenamefont {Mu}, \citenamefont {Song},
  \citenamefont {Jin}, \citenamefont {Liu},\ and\ \citenamefont
  {Wang}}]{xia2017micromagnetic}%
  \BibitemOpen
  \bibfield  {author} {\bibinfo {author} {\bibfnamefont {H.}~\bibnamefont
  {Xia}}, \bibinfo {author} {\bibfnamefont {Q.}~\bibnamefont {Zheng}}, \bibinfo
  {author} {\bibfnamefont {C.}~\bibnamefont {Mu}}, \bibinfo {author}
  {\bibfnamefont {C.}~\bibnamefont {Song}}, \bibinfo {author} {\bibfnamefont
  {C.}~\bibnamefont {Jin}}, \bibinfo {author} {\bibfnamefont {Q.}~\bibnamefont
  {Liu}}, \ and\ \bibinfo {author} {\bibfnamefont {J.}~\bibnamefont {Wang}},\
  }\href@noop {} {\bibfield  {journal} {\bibinfo  {journal} {Journal of
  Magnetism and Magnetic Materials}\ }\textbf {\bibinfo {volume} {432}},\
  \bibinfo {pages} {387} (\bibinfo {year} {2017})}\BibitemShut {NoStop}%
\bibitem [{\citenamefont {Yin}\ \emph {et~al.}(2014)\citenamefont {Yin},
  \citenamefont {Skomski}, \citenamefont {Sellmyer}, \citenamefont {Liou},
  \citenamefont {Russek}, \citenamefont {Evarts}, \citenamefont {Moreland},
  \citenamefont {Edelstein}, \citenamefont {Yuan}, \citenamefont {Yan} \emph
  {et~al.}}]{yin2014adjusting}%
  \BibitemOpen
  \bibfield  {author} {\bibinfo {author} {\bibfnamefont {X.}~\bibnamefont
  {Yin}}, \bibinfo {author} {\bibfnamefont {R.}~\bibnamefont {Skomski}},
  \bibinfo {author} {\bibfnamefont {D.}~\bibnamefont {Sellmyer}}, \bibinfo
  {author} {\bibfnamefont {S.-H.}\ \bibnamefont {Liou}}, \bibinfo {author}
  {\bibfnamefont {S.~E.}\ \bibnamefont {Russek}}, \bibinfo {author}
  {\bibfnamefont {E.~R.}\ \bibnamefont {Evarts}}, \bibinfo {author}
  {\bibfnamefont {J.}~\bibnamefont {Moreland}}, \bibinfo {author}
  {\bibfnamefont {A.}~\bibnamefont {Edelstein}}, \bibinfo {author}
  {\bibfnamefont {L.}~\bibnamefont {Yuan}}, \bibinfo {author} {\bibfnamefont
  {M.}~\bibnamefont {Yan}},  \emph {et~al.},\ }\href@noop {} {\bibfield
  {journal} {\bibinfo  {journal} {Journal of Applied Physics}\ }\textbf
  {\bibinfo {volume} {115}},\ \bibinfo {pages} {17E528} (\bibinfo {year}
  {2014})}\BibitemShut {NoStop}%
\bibitem [{\citenamefont {Donahue}\ and\ \citenamefont
  {Porter}(1999)}]{donahue1999oommf}%
  \BibitemOpen
  \bibfield  {author} {\bibinfo {author} {\bibfnamefont {M.~J.}\ \bibnamefont
  {Donahue}}\ and\ \bibinfo {author} {\bibfnamefont {D.~G.}\ \bibnamefont
  {Porter}},\ }\href@noop {} {\emph {\bibinfo {title} {OOMMF User's guide}}}\
  (\bibinfo  {publisher} {US Department of Commerce, Technology Administration,
  National Institute of Standards and Technology},\ \bibinfo {year}
  {1999})\BibitemShut {NoStop}%
\bibitem [{\citenamefont {Frankowski}\ \emph {et~al.}(2014)\citenamefont
  {Frankowski}, \citenamefont {Czapkiewicz}, \citenamefont {Skowro{\'n}ski},\
  and\ \citenamefont {Stobiecki}}]{frankowski2014micromagnetic}%
  \BibitemOpen
  \bibfield  {author} {\bibinfo {author} {\bibfnamefont {M.}~\bibnamefont
  {Frankowski}}, \bibinfo {author} {\bibfnamefont {M.}~\bibnamefont
  {Czapkiewicz}}, \bibinfo {author} {\bibfnamefont {W.}~\bibnamefont
  {Skowro{\'n}ski}}, \ and\ \bibinfo {author} {\bibfnamefont {T.}~\bibnamefont
  {Stobiecki}},\ }\href@noop {} {\bibfield  {journal} {\bibinfo  {journal}
  {Physica B: Condensed Matter}\ }\textbf {\bibinfo {volume} {435}},\ \bibinfo
  {pages} {105} (\bibinfo {year} {2014})}\BibitemShut {NoStop}%
\bibitem [{\citenamefont {Ch{\k{e}}ci{\'n}ski}\ and\ \citenamefont
  {Frankowski}(2016)}]{chkecinski2016mage}%
  \BibitemOpen
  \bibfield  {author} {\bibinfo {author} {\bibfnamefont {J.}~\bibnamefont
  {Ch{\k{e}}ci{\'n}ski}}\ and\ \bibinfo {author} {\bibfnamefont
  {M.}~\bibnamefont {Frankowski}},\ }\href@noop {} {\bibfield  {journal}
  {\bibinfo  {journal} {Computer Physics Communications}\ }\textbf {\bibinfo
  {volume} {207}},\ \bibinfo {pages} {487} (\bibinfo {year}
  {2016})}\BibitemShut {NoStop}%
\bibitem [{\citenamefont {Landau}\ and\ \citenamefont
  {Lifshitz}(1935)}]{landau1935theory}%
  \BibitemOpen
  \bibfield  {author} {\bibinfo {author} {\bibfnamefont {L.~D.}\ \bibnamefont
  {Landau}}\ and\ \bibinfo {author} {\bibfnamefont {E.}~\bibnamefont
  {Lifshitz}},\ }\href@noop {} {\bibfield  {journal} {\bibinfo  {journal}
  {Phys. Z. Sowjetunion}\ }\textbf {\bibinfo {volume} {8}},\ \bibinfo {pages}
  {101} (\bibinfo {year} {1935})}\BibitemShut {NoStop}%
\bibitem [{\citenamefont {Gilbert}(1955)}]{gilbert1955lagrangian}%
  \BibitemOpen
  \bibfield  {author} {\bibinfo {author} {\bibfnamefont {T.}~\bibnamefont
  {Gilbert}},\ }\href@noop {} {\bibfield  {journal} {\bibinfo  {journal} {Phys.
  Rev.}\ }\textbf {\bibinfo {volume} {100}},\ \bibinfo {pages} {1243} (\bibinfo
  {year} {1955})}\BibitemShut {NoStop}%
\bibitem [{\citenamefont {Gilbert}(2004)}]{gilbert2004phenomenological}%
  \BibitemOpen
  \bibfield  {author} {\bibinfo {author} {\bibfnamefont {T.~L.}\ \bibnamefont
  {Gilbert}},\ }\href@noop {} {\bibfield  {journal} {\bibinfo  {journal}
  {Magnetics, IEEE Transactions on}\ }\textbf {\bibinfo {volume} {40}},\
  \bibinfo {pages} {3443} (\bibinfo {year} {2004})}\BibitemShut {NoStop}%
\bibitem [{\citenamefont {Slonczewski}(1996)}]{slonczewski1996current}%
  \BibitemOpen
  \bibfield  {author} {\bibinfo {author} {\bibfnamefont {J.~C.}\ \bibnamefont
  {Slonczewski}},\ }\href@noop {} {\bibfield  {journal} {\bibinfo  {journal}
  {Journal of Magnetism and Magnetic Materials}\ }\textbf {\bibinfo {volume}
  {159}},\ \bibinfo {pages} {L1} (\bibinfo {year} {1996})}\BibitemShut
  {NoStop}%
\bibitem [{\citenamefont {Slavin}\ and\ \citenamefont
  {Tiberkevich}(2009)}]{slavin2009nonlinear}%
  \BibitemOpen
  \bibfield  {author} {\bibinfo {author} {\bibfnamefont {A.}~\bibnamefont
  {Slavin}}\ and\ \bibinfo {author} {\bibfnamefont {V.}~\bibnamefont
  {Tiberkevich}},\ }\href@noop {} {\bibfield  {journal} {\bibinfo  {journal}
  {IEEE Transactions on Magnetics}\ }\textbf {\bibinfo {volume} {45}},\
  \bibinfo {pages} {1875} (\bibinfo {year} {2009})}\BibitemShut {NoStop}%
\end{thebibliography}%

\end{document}